\documentclass[preprint,authoryear,12pt]{elsarticle}

\usepackage{epsfig}
\usepackage{rotating}

\usepackage{amssymb}
\usepackage{graphicx, amsmath}
\usepackage{amsthm,amssymb}
\usepackage{multirow}
\usepackage{longtable}
\usepackage{cite}
\usepackage{amsfonts}
\usepackage{subfig}

\usepackage[ps2pdf,%
a4paper=true,%
breaklinks=true,%
colorlinks=true,%
pdfauthor={First Author et al.},%
pdftitle={Template for manuscripts in Advances in Space Research}%
]{hyperref}

\journal{Advances in Space Research}

\begin{document}

\begin{frontmatter}



\title{Exposure Time Calculator for Immersion Grating Infrared Spectrograph: IGRINS}


\author[khu]{Huynh Anh N. Le}\ead{huynhanh7@khu.ac.kr}
\author[khu]{Soojong Pak\corref{cor}}\ead{soojong@khu.ac.kr}
\author[ut]{Daniel T. Jaffe}
\author[ut]{Kyle Kaplan}
\author[kasi]{Jae-Joon Lee}
\author[snu]{Myungshin Im}
\author[chi]{Andreas Seifahrt}

\address[khu]{School of Space Research and Institute of Natural Sciences, Kyung Hee University, \\
 1732 Deogyeong-daero, Giheung-gu, Yongin-si, Gyeonggi-do, 446-701, Republic of Korea}
\address[ut]{Department of Astronomy, University of Texas at Austin, Austin, Texas, TX 78712, USA}
\address[kasi]{Korea Astronomy and Space Science Institute, Daejeon, 305-348, Republic of Korea}
\address[snu]{Department of Physics and Astronomy, Seoul National University, \\
Center for the Exploration of the Origin of the Universe (CEOU), Seoul, 151-742, Republic of Korea}
\address[chi]{Department of Astronomy and Astrophysics, University of Chicago, Chicago, IL 60637, USA}


\cortext[cor]{Corresponding author}

\begin{abstract}

We present an exposure-time calculator (ETC) for the Immersion Grating Infrared Spectrograph (IGRINS). The signal and noise values are calculated by taking into account the telluric background emission and absorption, the emission and transmission of the telescope and instrument optics, and the dark current and read noise of the infrared detector arrays. For the atmospheric transmission, we apply models based on the amount of precipitable water vapor along the line of sight to the target. The ETC produces the expected signal-to-noise ratio (S/N) for each resolution element, given the exposure-time and number of exposures. In this paper, we compare the simulated continuum S/N for the early-type star HD~124683 and the late-type star GSS~32, and the simulated emission line S/N for the H$_{2}$ rovibrational transitions from the Iris Nebula NGC~7023 with the observed IGRINS spectra. The simulated S/N from the ETC is overestimated by 40$-$50$\%$ for the sample continuum targets.

\end{abstract}


\end{frontmatter}

\parindent=0.5 cm


\section{INTRODUCTION}

\begin{sidewaysfigure}[!thp]
\centering
    \includegraphics[width=120ex]{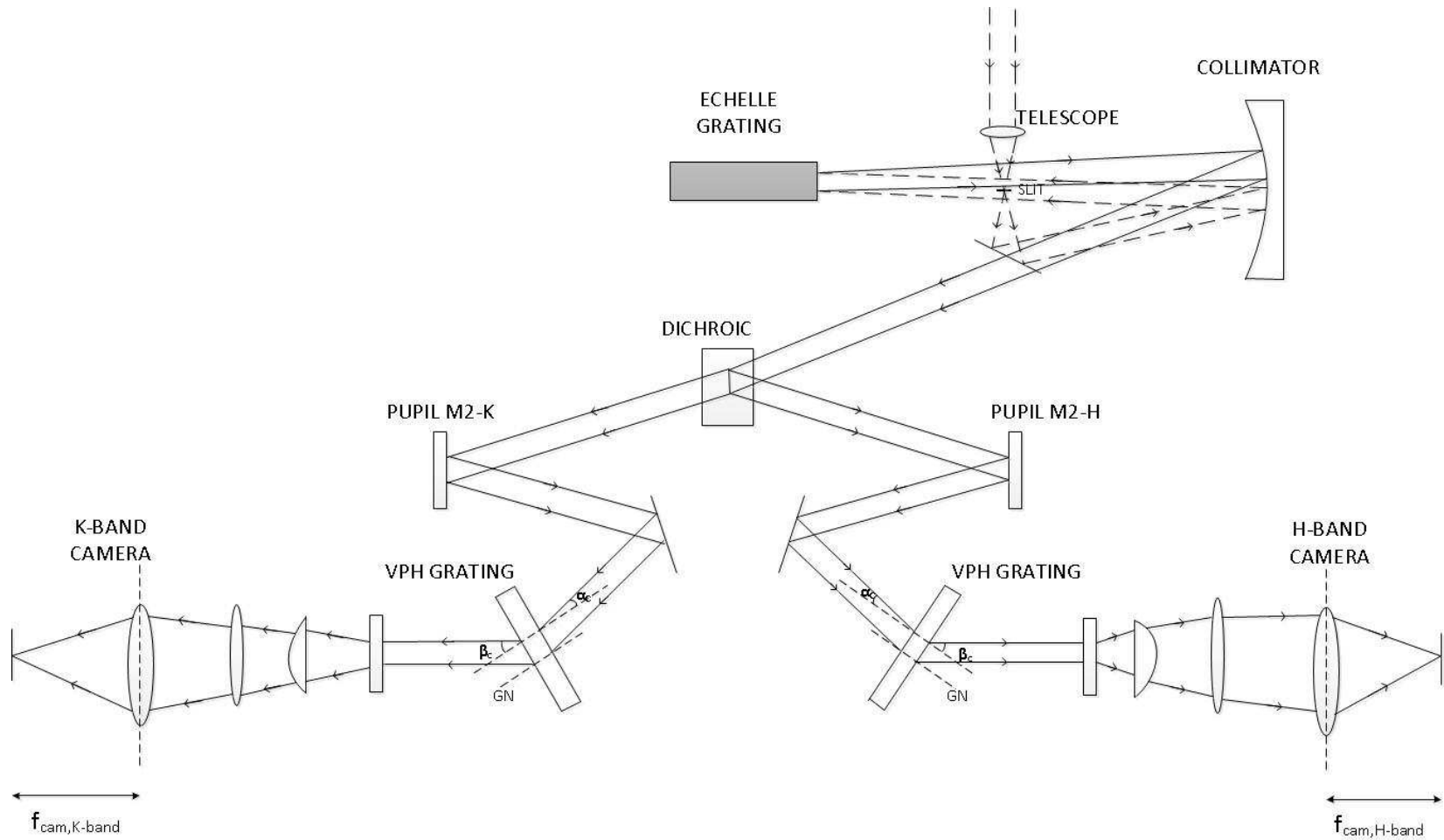}
\caption{IGRINS optical design sketch.}
\label{sketch}
\end{sidewaysfigure}

The Immersion Grating Infrared Spectrograph (IGRINS) is a cross dispersed near-infrared spectrograph designed using silicon immersion echelle grating (e.g., \citealp{Marsh07}; \citealp{Wang10}; \citealp{Santiago12}). This instrument can cover the whole $H$ and $K$ bands ($1.4$ $-$ $2.5$ $\mu$m) in single exposure with a resolving power of $R$ $=$ 40,000. Two volume phase holographic (VPH) gratings are used as cross-dispersing elements in the $H$ and $K$ bands. IGRINS employed two 2048 $\times$ 2048 pixel Teledyne Scientific \& Imaging HAWAII-2RG detectors with SIDECAR ASIC cryogenic controllers. Figure \ref{sketch} shows the optical design sketch of IGRINS. This instrument has been employed on the $2.7$ m Harlan J. Smith Telescope at the McDonald Observatory since 2014 March. Detailed descriptions of IGRINS can be found in \citet{Yuk10} and \citet{Park14}.

Exposure-time calculators (ETCs) play an important role in preparing for observations by accurately simulating the expected exposure-time to achieve a required signal-to-noise ratio (S/N). Examples of ETCs include those for MOSFIRE\footnote{http://www2.keck.hawaii.edu/inst/mosfire/etc.html.}, CRIRES$^{2}$, SOFI$^{2}$, KMOS\footnote{See the ESO exposure-time calculators at http://www.eso.org/observing/etc.}, HSpot\footnote{http://herschel.asdc.asi.it/index.php?page=proposal.html.}, and the Wide Field Planetary Camera 2 (WFPC2) of the Hubble Space Telescope (e.g., \citealp{Bernstein02}; \citealp{Causi02}).

We developed the IGRINS ETC by taking into account the telluric background emission and absorption, the thermal emission and throughput of the telescope and instrument optics, and the dark current and read noise of the infrared detector arrays. This tool is a part of the IGRINS software which includes the IGRINS data reduction pipeline package \citep{Sim14}. The software was written in the interactive data language Python, and the user controls specific calculation modes and input parameters via a graphical user interface. In this paper, we present detailed calculation algorithms and the results of the IGRINS ETC.


\section{SENSITIVITY SIMULATION METHODS}

\subsection{Throughput}

We define the total throughput, $\tau_{\mathrm{point}}$ for point sources as

\begin{equation}\label{eq:tau}
\tau _{\mathrm{point}} = \tau _{\mathrm{atmo}} \tau _{\mathrm{mirror}} \tau_{\mathrm{gr}} \tau _{\mathrm{vph}} \tau _{\mathrm{optics}} \tau _{\mathrm{slit}} \tau _{\mathrm{qe}}
\end{equation}

\begin{equation}\label{eq:tau_ex}
\tau _{\mathrm{extend}} = \tau _{\mathrm{atmo}} \tau _{\mathrm{mirror}} \tau_{\mathrm{gr}} \tau _{\mathrm{vph}} \tau _{\mathrm{optics}}  \tau _{\mathrm{qe}}
\end{equation}

where $\tau$ terms are the throughput values of the average atmosphere ($\tau _{\mathrm{atmo}}$), telescope mirrors ($\tau _{\mathrm{mirror}}$), immersion grating ($\tau_{\mathrm{gr}}$), VPH gratings ($\tau_{\mathrm{vph}}$), other IGRINS optics ($\tau _{\mathrm{optics}}$), slit ($\tau _{\mathrm{slit}}$), and detector ($\tau _{\mathrm{qe}}$). Table \ref{tau_emissivity} shows the estimated values of these parameters in equations (\ref{eq:tau}) and (\ref{eq:tau_ex}). We estimate that the average throughput value of the atmosphere is 0.77 and 0.83, assuming that the precipitable water vapor (PWV) is 4 mm, at the wavelength ranges of $1.4$$-$$1.9$ $\mu$m and $1.9$$-$$2.5$ $\mu$m for the $H$ and $K$ bands, respectively. The model of the atmosphere transmission is shown in Appendix A. Since the point spread function (PSF) on the slit closely resembles a Gaussian, we can easily estimate the slit loss for a given image size. We define a slit throughput, $\tau_{\mathrm{slit}}$, as

\begin{equation}
\tau_{\mathrm{slit}} = \frac{F_{\mathrm{slit}} }{F_{\circ}}
\end{equation}

where $F_{\circ}$ is the unobscured flux from the target, and $F_{\mathrm{slit}}$ is the flux passing through the slit \citep{Lee06}. If the PSF is a circularly symmetric Gaussian function, $\tau_{\mathrm{slit}}$ can be expressed as the error function:

\begin{equation}\label{eq:cslit}
\tau _{\mathrm{slit}} = \frac{1}{\sqrt{2\pi }\sigma }\int_{-Y}^{+Y}\exp\left [ -\frac{y^{2}}{2\sigma ^{2}} \right ]dy
\end{equation}

where $Y$ is a half of the slit width and $\sigma$ is a Gaussian width, i.e., $\sigma$ = $\Delta y_{\mathrm{FWHM}}$ / ($2\sqrt{2\ln 2}$), where $\Delta y_{\mathrm{FWHM}}$ is the measured full width at half maximum along the slit length.

\subsection{Background Emissions}

We calculate the background emissions from the Zodiacal light, the Moon light, OH emission lines, and thermal emission from the atmosphere. Table \ref{background} shows the parameters used in the calculations. The number of electrons s$^{-1}$ pixel$^{-1}$ by the background emissions is given by

\begin{equation}\label{eq:zod}
\varepsilon_{\mathrm{Zod,Moon,OH}} = A\Omega \tau_{\mathrm{extend}} \frac{\lambda _{\circ}}{\Delta \lambda_{\mathrm{BW}} R}\left ( \Phi _{\mathrm{Zod}}+ \Phi _{\mathrm{Moon}} +0.1\Phi _{\mathrm{OH}} \right )
\end{equation}

where $\tau_{\mathrm{extend}}$ is the total throughput for an extended source, $R$ is the spectral resolution ($R$ $=$ 40,000) for IGRINS, $\lambda _{\circ}$ is the central wavelength of the $H$ or $K$ filter, $\Delta \lambda_{\mathrm{BW}}$ is the band width of the filter, $\Phi _{\mathrm{Zod}}$, $\Phi _{\mathrm{Moon}}$, and $\Phi _{\mathrm{OH}}$ are photon fluxes from the Zodiacal scattered light, the Moon light, and OH airglow in units of photon s$^{-1}$ m$^{-2}$ arcsec$^{-2}$, $A\Omega$ is the aperture of the telescope multiplied by the solid angle on the sky viewed by one pixel ($A\Omega$ $=$ 1.00 $\times$ 10$^{-11}$ m$^{2}$ sr pixel$^{-1}$).

In Table \ref{background}, we list the average values of the Zodiacal light\footnote{See the Sky Model Calculator at http://www.eso.org/observing/etc.} at ecliptic latitudes of 0$^{\circ}$, 30$^{\circ}$, 60$^{\circ}$, and 90$^{\circ}$. In the simulation, we assume the heliocentric ecliptic longitudes at $\pm$90$^{\circ}$ since the Zodiacal light is not variable at different ecliptic longitudes. In the ETC, the average values of the Zodiacal light is $\Phi _{\mathrm{Zod}}$ = 55, 41 photon s$^{-1}$ m$^{-2}$ arcsec$^{-2}$ for $H$ and $K$ bands, respectively.

\begin{figure*}[!thp]
\centering
    \includegraphics[width=75ex]{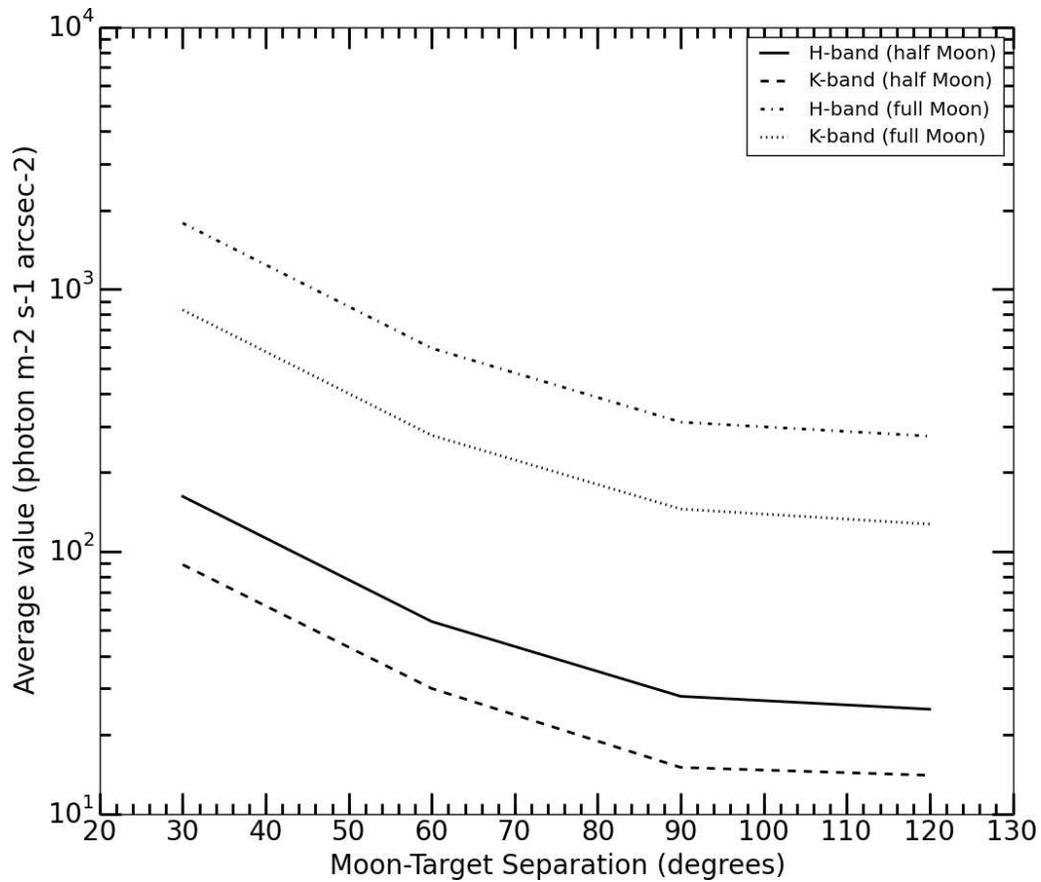}
\caption{Average values of the Moon light at different Moon-target separations in the case of the half Moon phase (solid and dashed lines) and in the case of the full Moon phase (dash-dot and dotted lines) in $H$ and $K$ bands, respectively.}
\label{moon}
\end{figure*}

The Moon light is another natural source that is the major contributor to the sky background (e.g., \citealp{Noll12}; \citealp{Jones13}). We calculate the average values of the Moon light$^{4}$ at the Moon phases of the half Moon and the full Moon for the cases of the Moon and target separations at 0$^{\circ}$, 30$^{\circ}$, 60$^{\circ}$, 90$^{\circ}$, and 120$^{\circ}$ (See Figure \ref{moon}). In the ETC, we assume that the target separation is 60$^{\circ}$ for the cases of the half Moon and the full Moon phases. We may consider to apply for all the cases of the Moon light in the next versions of the ETC.

\begin{figure*}[!h]
\centering
    \includegraphics[width=50ex,keepaspectratio]{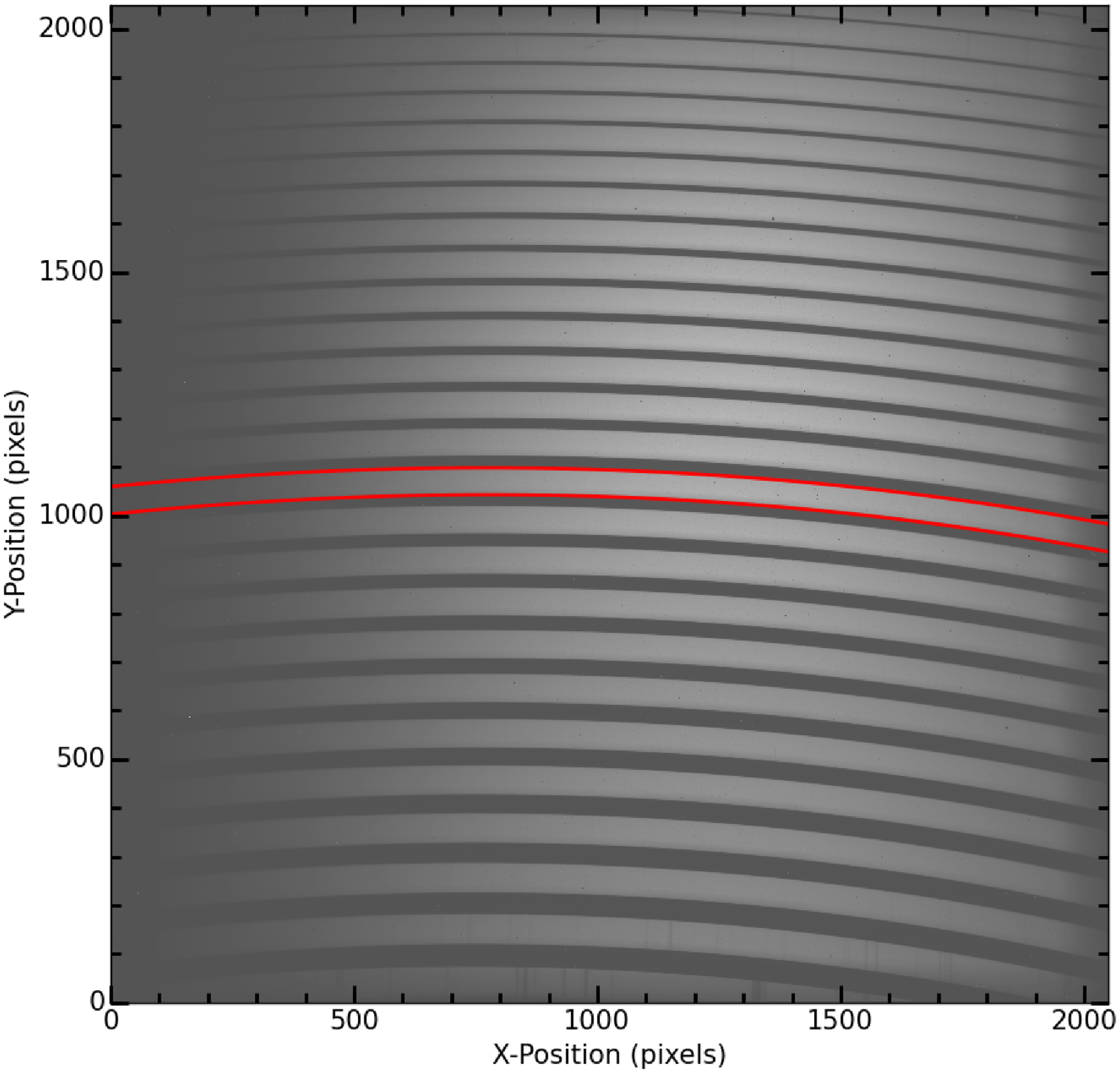}
    \includegraphics[width=50ex,keepaspectratio]{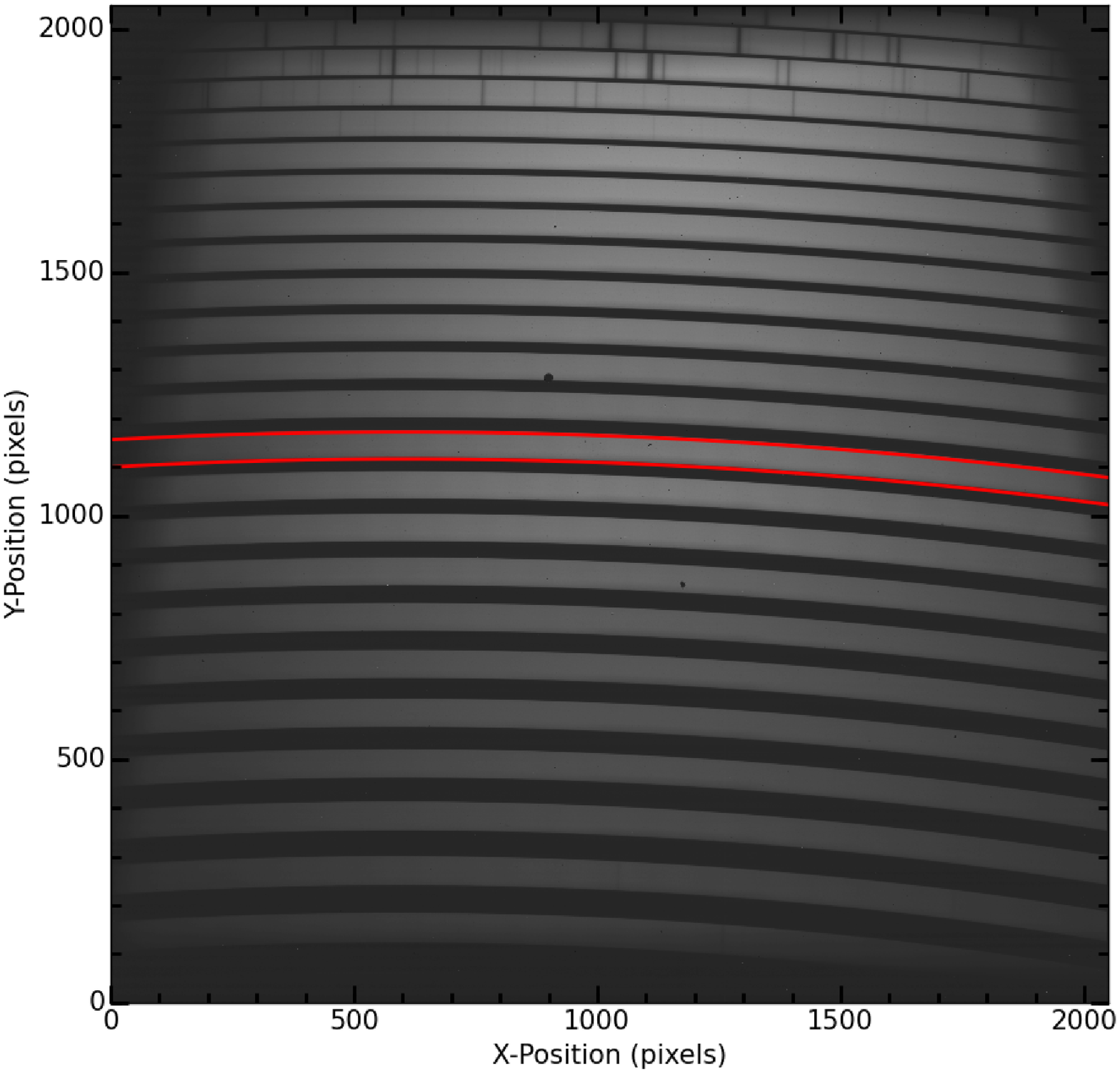}
\caption{Flat image in the $H$ band (top plot) and $K$ band (bottom plot). The red lines show the example of an order region in the echellogram. The inter-order regions are the remaining parts of the image. }
\label{flat}
\end{figure*}

We calculate the $\Phi _{\mathrm{OH}}$ value based on OH emission lines which are shown in Figure \ref{ohlines} in Appendix A. We sum up the OH line intensity values of whole band. The OH value, used in equation (\ref{eq:zod}) is the summation of OH line intensity in $H$ and $K$ bands, $\Phi _{\mathrm{OH}}$ = 3390 + 1096 = 4486 photon s$^{-1}$ m$^{-2}$ arcsec$^{-2}$. We also consider that these is ratio of photons from OH lines that are scattered inside the IGRINS optical box and distributed as scattered light across the detector. By using Flat image which is subtracted by Dark image (See Figure \ref{flat}), we calculate the average intensity of pixel values in the order regions, and the inter-order regions, then estimate the scattering ratio of photon from OH lines inside the optics (See Table \ref{order}). The calculated ratio value is 0.1 for both $H$ and $K$ bands.

The background thermal flux in units of electrons s$^{-1}$ pixel$^{-1}$ is calculated by

\begin{equation}\label{eq:thermal}
\varepsilon_{\mathrm{thermal}} = A\Omega \varepsilon \frac{B_{\nu,T }}{hR}
\end{equation}

where $h$ is the Planck constant, and $B_{\nu,T}$ is the Planck function of frequency $\nu$ and temperature $T$ in units of W m$^{-2}$ Hz$^{-1}$ sr$^{-1}$. We define the total emissivity $\varepsilon$ as

\begin{equation}\label{eq:emissivity}
\varepsilon = \frac{\left ( 1 - \tau _{\mathrm{atmo}} \right ) + \left [ \varepsilon_{\mathrm{mirror}} + \left ( 1-\tau _{\mathrm{window}} \right ) / \tau _{\mathrm{window}}\right ]/\tau _{\mathrm{mirror}}}{\tau _{\mathrm{atmo}}} \tau _{\mathrm{point}} \tau _{\mathrm{slit}}
\end{equation}

where $\varepsilon_{\mathrm{mirror}}$ is 0.25, $\tau_{\mathrm{mirror}}$ is 0.55, and $\tau_{\mathrm{window}}$ is 0.95. \\

\subsection{Signal-to-Noise for Continuum}

We observe point sources using the ``Nod-on-Slit'' mode, where we take spectra at two positions along the slit and subtract them from each other \citep{Sim14}. This subtracts various sources of background while simultaneously gathering signal from the science target during all exposures.

The continuum signal is
\begin{equation}\label{eq:signal}
S_{\mathrm{Cont}} = \frac{n_{\mathrm{exp}}t_{\mathrm{exp}}A\tau_{\mathrm{point}} F_{\mathrm{ZM}}10^{-0.4m}}{hR}
\end{equation}

where $S_{\mathrm{Cont}}$ is in units of electrons per resolution element, $F_{\mathrm{ZM}}$ is the zero magnitude flux density in units of W m$^{-2}$ Hz$^{-1}$, $m$ is the apparent magnitude of the continuum source, $n_{\mathrm{exp}}$ is the number of exposures, $t_{\mathrm{exp}}$ is the exposure-time, and $\tau_{\mathrm{point}}$ is the total throughput for the point source.

The electrons collected by the detector exhibit as a Poisson distribution. The noise is calculated from the formula as

\begin{equation}\label{eq:mnoise}
N_{\mathrm{Cont}} =  \left [\mathrm{Q} + S_{\mathrm{Cont}} \right ]^{1/2}
\end{equation}

where $N_{\mathrm{Cont}}$ is in units of electrons per resolution element, and

\begin{equation}\label{eq:mnoise_N}
\mathrm{Q} = 2p_{\mathrm{slit}}^{2}n_{\mathrm{exp}} \left [t_{\mathrm{exp}}\left (\varepsilon_{\mathrm{Zod,Moon,OH}}+ d_{\mathrm{dark}} + \varepsilon_{\mathrm{thermal}}  \right ) + d_{\mathrm{read}}^{2} \right ]
\end{equation}

In this formula, $p_{\mathrm{slit}}$ is 3.66 pixels in the spatial and spectral resolutions, $d_{\mathrm{dark}}$ and $d_{\mathrm{read}}$ are the dark current and read noise of the infrared detector array, $\varepsilon_{\mathrm{Zod,Moon,OH}}$ and $\varepsilon_{\mathrm{thermal}}$ are given by equations in (\ref{eq:zod}) and (\ref{eq:thermal}). For more realistic simulations, we should have considered the number of pixels in spatial and spectral resolution elements instead of $p_{\mathrm{slit}}^{2}$ pixels. Since we always subtract a background frame, the noise term has a factor of 2 in equation (\ref{eq:mnoise_N}).

\subsection{Signal and Noise for Emission Lines}
The ETC can simulate the IGRINS observations of emission lines given a rest wavelength, flux, doppler shift and width. We use the following definitions for the signal and noise of the emission line \citep{Pak04}:

\begin{equation}\label{eq:signalline}
S_{\mathrm{Line}} = \Delta \lambda_{\mathrm{pixel}} \sum_{\mathrm{i}}\left (  f_{\mathrm{L,i}} - \bar{f_{\mathrm{C}}}  \right )
\end{equation}

\begin{equation}\label{eq:noiseline}
N_{\mathrm{Line}} = \Delta \lambda_{\mathrm{pixel}} \sigma (f_{\mathrm{C,i}})\sqrt{n_{\mathrm{L}}\left ( 1+1/n_{\mathrm{C}} \right )}
\end{equation}

where $f_{\mathrm{L,i}}$ is the line flux data samples in units of W m$^{-2}$ $\mu$m$^{-1}$, $\bar{f_{\mathrm{C}}}$ is the average continuum adjacent to the emission line, $\Delta \lambda_{\mathrm{pixel}}$ is the wavelength range corresponding to the pixel width at line center, $\sigma (f_{\mathrm{C,i}})$ is standard deviation of the continuum, and $n_{\mathrm{L}}$ and $n_{\mathrm{C}}$ are the sample numbers in the emission line and continuum bands, respectively.

\section{SIMULATION RESULTS}

\subsection{PHOTOMETRIC MODE}

In the photometric mode, the ETC returns the average S/N for the $H$ and $K$ bands, assuming that the target is a point source. In this mode, the S/N values are computed for whole spectral $H$ and $K$ bands.
The users can input the precipitable water vapor (PWV), the exposure-time, the number of exposures, the expected $K$ band magnitude, the effective temperature of the source, the seeing size, and the Moon light phases.

\subsection{SPECTRAL MODE}

\begin{figure*}[!thp]
\centering
\includegraphics[width=75ex]{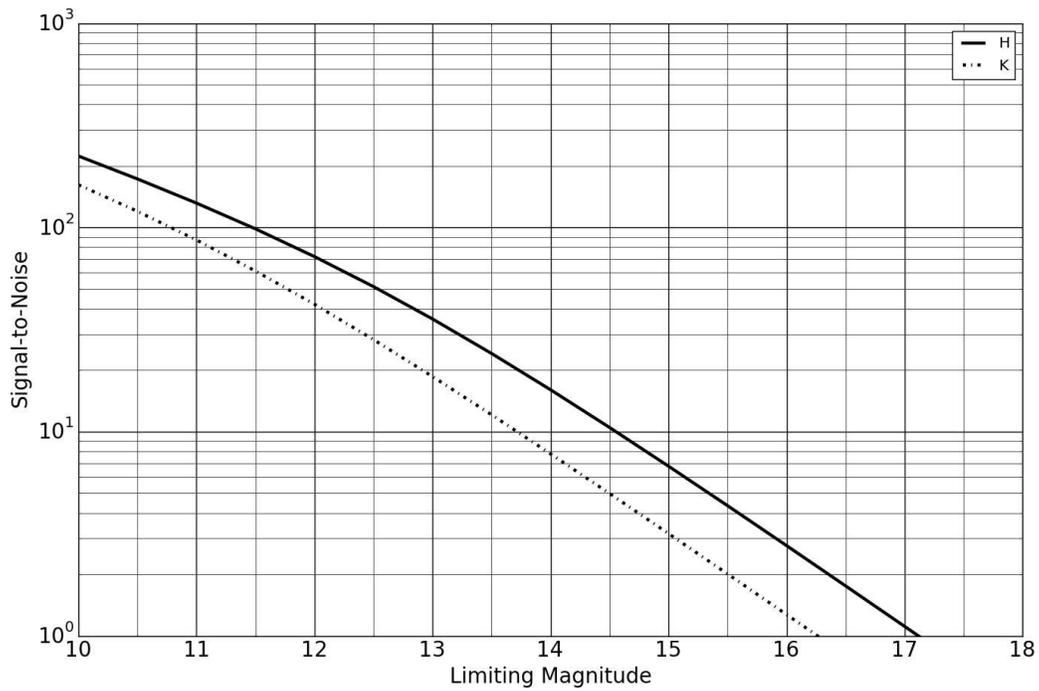}
\caption{Plot of S/N vs. limiting magnitude for continuum point sources in the $H$ band and $K$ band with exposure-time per frame, t = $600$ s, and number of exposures, N = $6$. The seeing is 1.2 arcsec and the PWV is 2 mm. The Moon light is not included in the calculation.}
\label{signal_magnitude}
\end{figure*}

\begin{figure*}[!thp]
\centering
    \includegraphics[width=61.5ex,height=60ex,keepaspectratio]{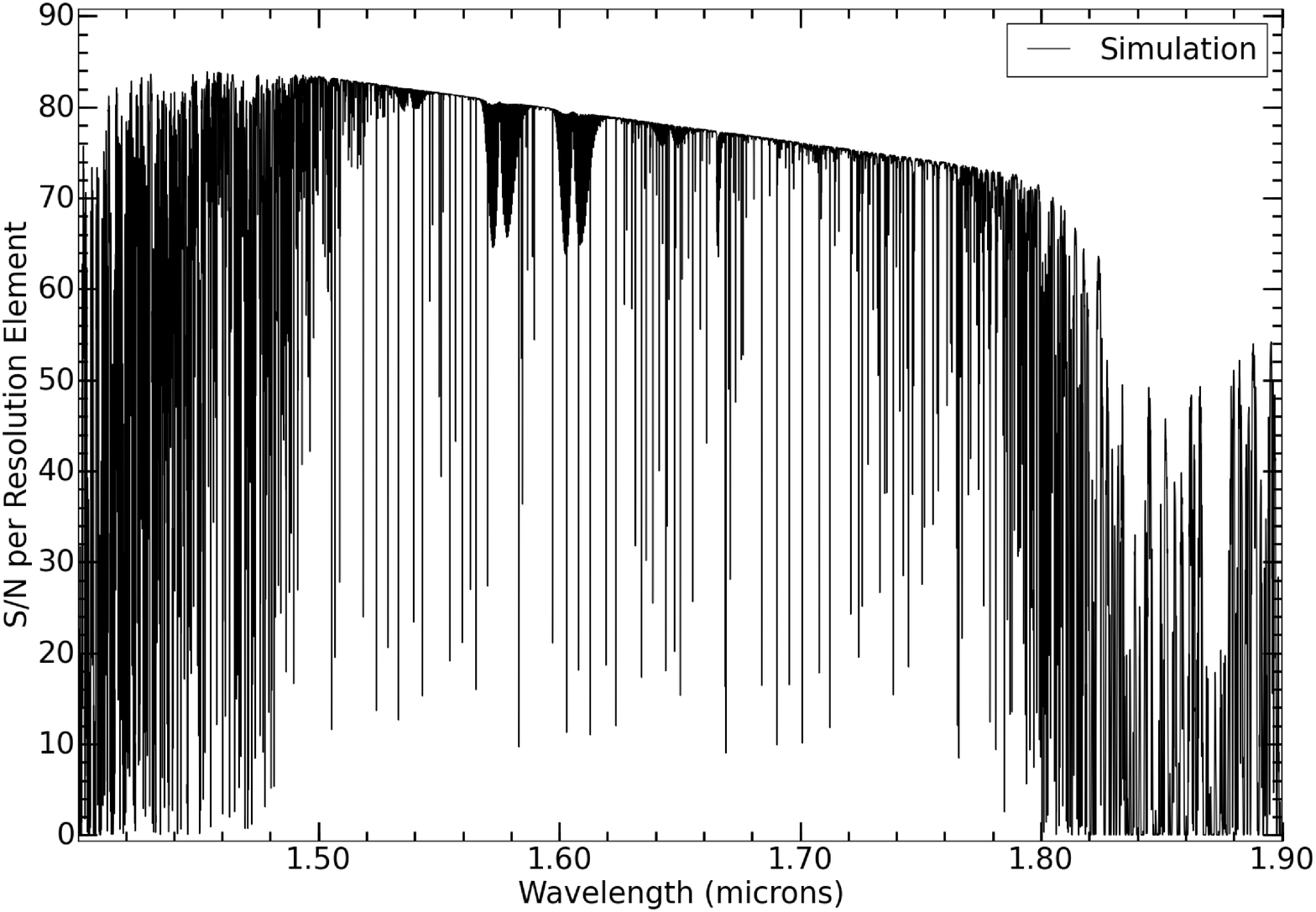}
    \includegraphics[width=60ex,height=60ex,keepaspectratio]{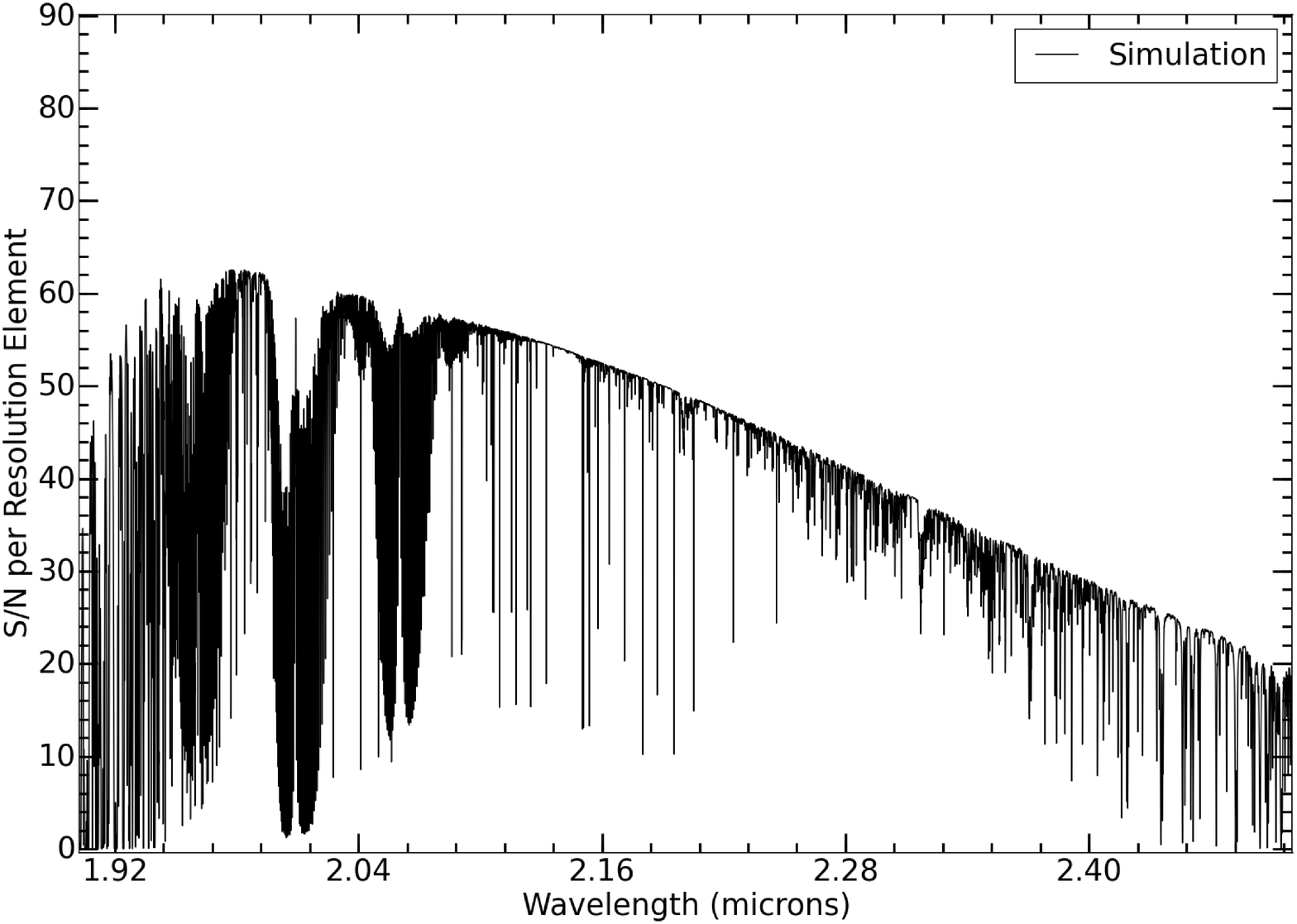}
\caption{Plots of S/N vs. wavelength in the $H$ band and $K$ band. We show a point blackbody (T = 6000 K) source with 12.1 mag in the $H$ band and 12.0 mag in the $K$ band. The seeing is 1.2 arcsec and the PWV is 2 mm. The exposure-time per frame is $600$ s, and the number of exposures is $6$. The Moon light is not included in the calculation.}
\label{signal_wavelength}
\end{figure*}

The spectral mode allows the users to view plots of the calculated S/N as a function of magnitude or wavelength. Figure \ref{signal_magnitude} shows the plot of S/N vs. limiting magnitude for continuum point sources in the $H$ and $K$ bands. Figure \ref{signal_wavelength} shows the plots of S/N vs. wavelength in the $H$ and $K$ bands for a blackbody source with a temperature of 6000 K, H-magnitude of 12.1, and K-magnitude of 12.0. In this spectral mode, the S/N values are variable with wavelength. In this mode, the throughput values of the atmosphere and the background signal and noise from OH emission lines are depended on wavelength. The model atmosphere is shown in Appendix A. Figures \ref{atmospheric_hband}, \ref{atmospheric_kband}, and \ref{ohlines} show the atmospheric transmission and OH emission lines in the $H$ and $K$ bands.

\section{TEST OBSERVATIONS}

IGRINS had three commissioning runs, in 2014 March, May and July, on the 2.7 m Harlan J. Smith Telescope at McDonald Observatory. In order to compare our ETC simulation results with real observational results, we chose an early-type A0V star, HD~124683, and a late-type K5 star, GSS~32, for point source continuum targets, and a reflection nebulae, NGC~7023 for an extended line emission target. The data of HD~124683 and GSS~32 were taken by using the ``Nod-on-Slit'' mode on 2014 May 27. The total exposure-times are 480~s and 960~s, respectively. NGC~7023 was observed by using the ``Nod-off-Slit'' mode on 2014 July 13. The total exposure-time is 1200~s.

NGC~7023 is a typical example of a photodissociation region (\citealp{Lemaire96}; \citealp{Lemaire99}). The spectra of this target show many strong narrow emission lines from the molecular rovibrational transitions of H$_{2}$ (e.g., \citealp{Martini97}; \citealp{Martini99}). The observed H$_{2}$ emission lines from this target are good examples to test the simulated emission lines from the ETC. We use the A0V standard star, HD~155379, to do the absolute flux calibration for NGC~7023 (see Table \ref{standard}). The flux calibration processes are based on the method of \citet{Lee06}.

\subsection{Signal-to-Noise for Continuum Source}

\begin{figure*}[!thp]
\centering
    \includegraphics[width=60ex,height=60ex,keepaspectratio]{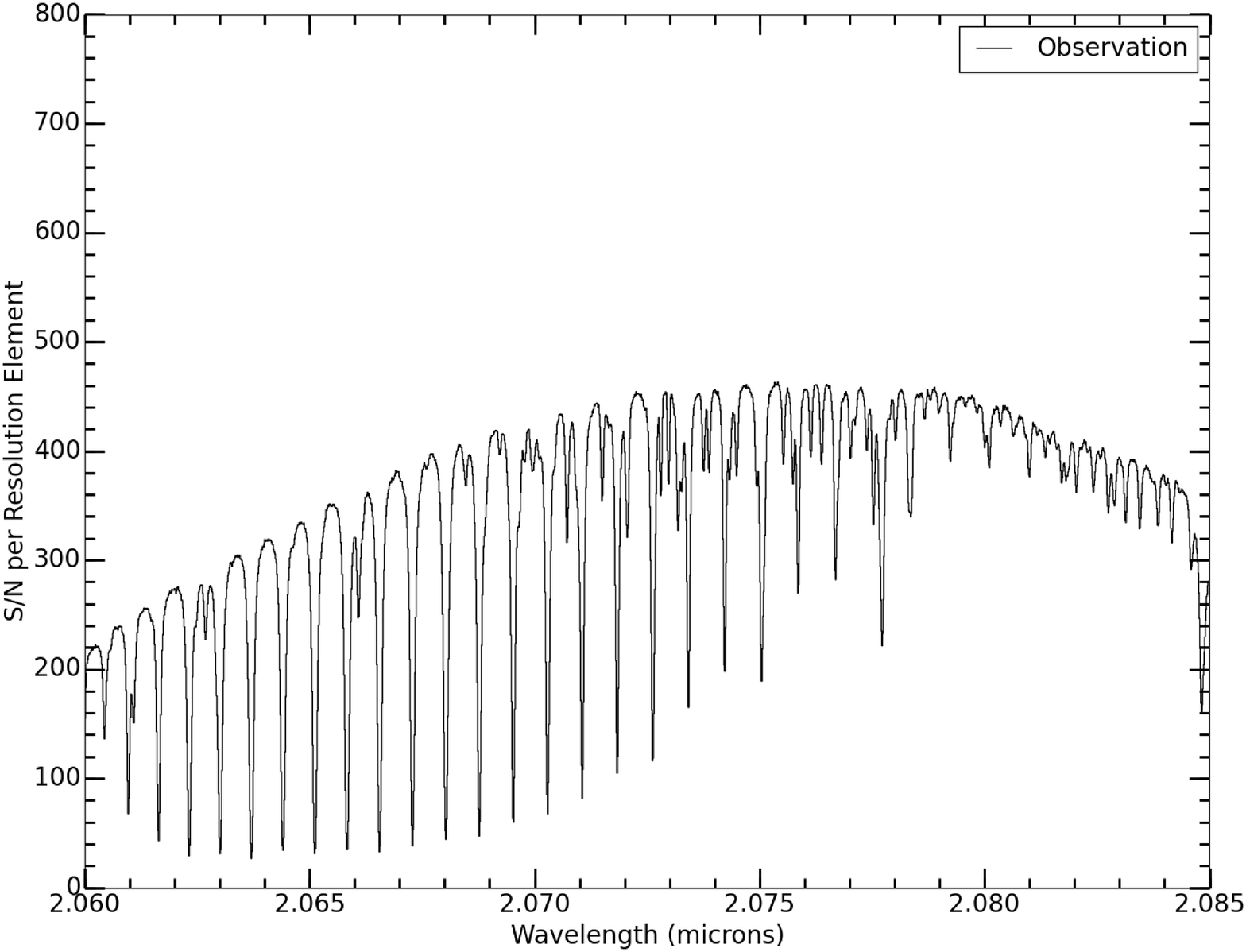}
    \includegraphics[width=60ex,height=60ex,keepaspectratio]{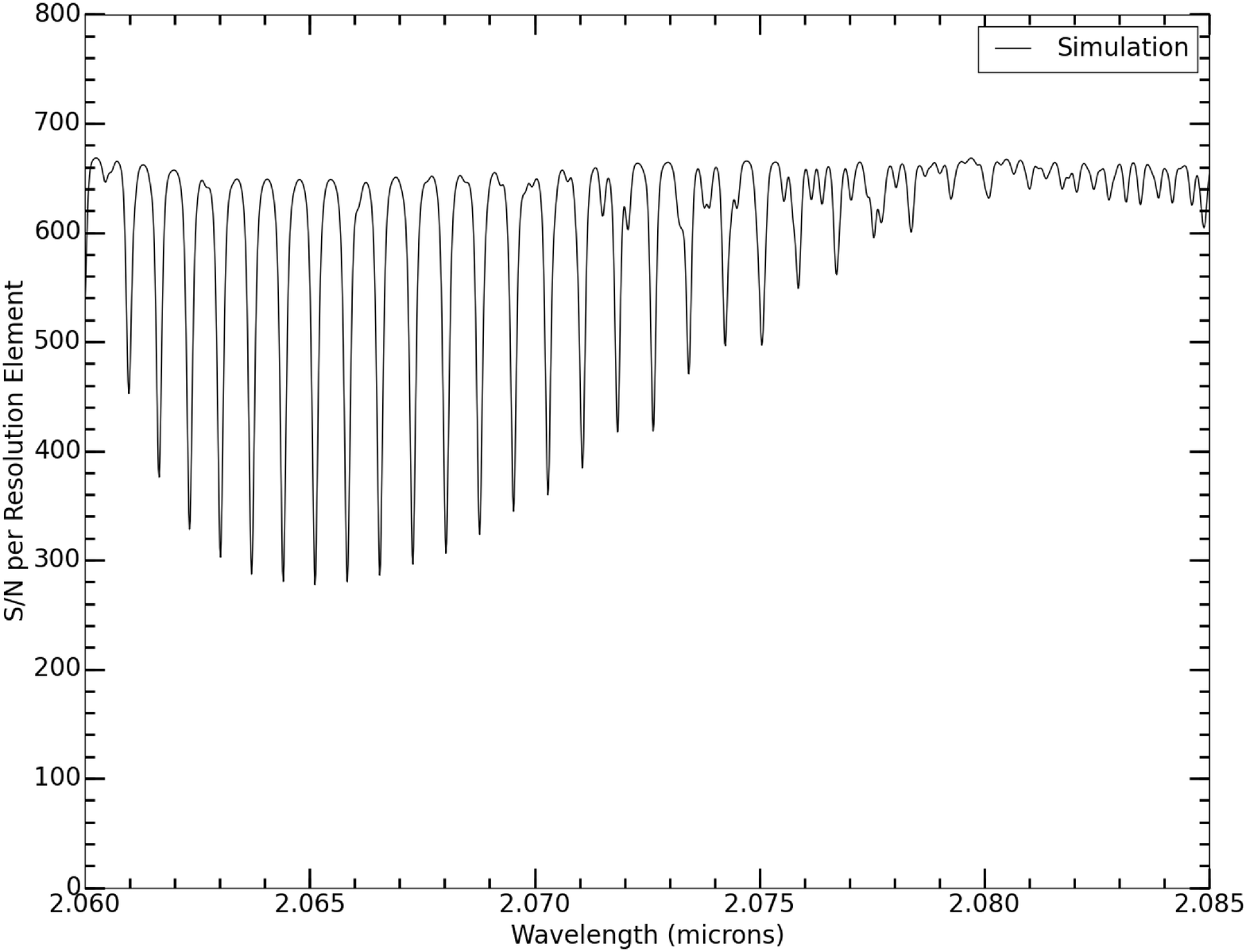}
\caption{Plots of S/N vs. wavelength of HD~124683 from the observational data (top plot) and simulation data from the ETC (bottom plot). The simulated data from the ETC assumes exposure times of  t = 120 s, number of exposures  N = 4, the K-magnitude m$_{K}$ = 5.552 mag, and the effective temperature of the source T = 9500 K. The seeing is 0.9 arcsec. The PWV is assumed to be 2 mm. The Moon light is not included in the calculation.}
\label{aov_sn}
\end{figure*}

\begin{figure*}[!thp]
\centering
    \includegraphics[width=60ex,height=60ex,keepaspectratio]{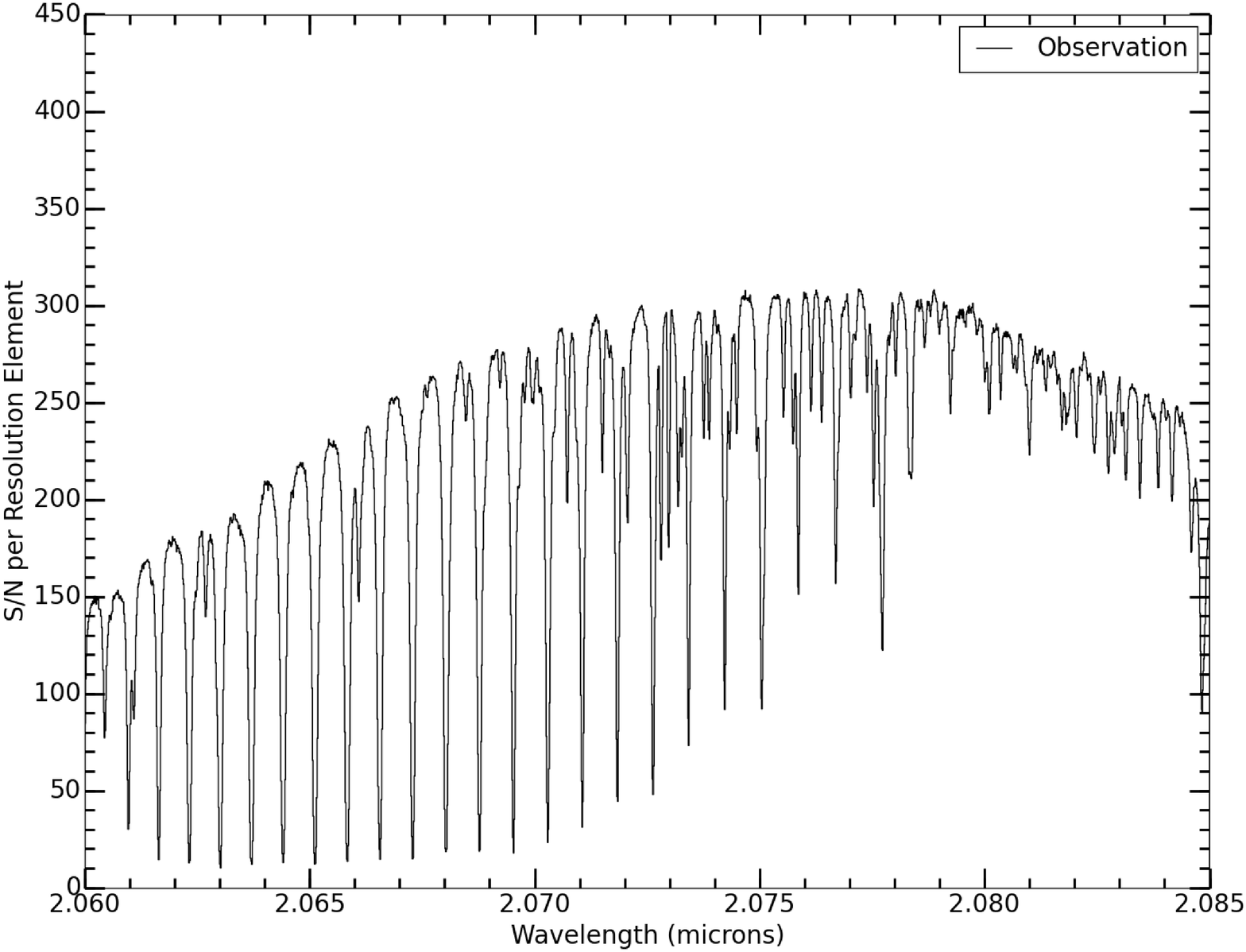}
    \includegraphics[width=60ex,height=60ex,keepaspectratio]{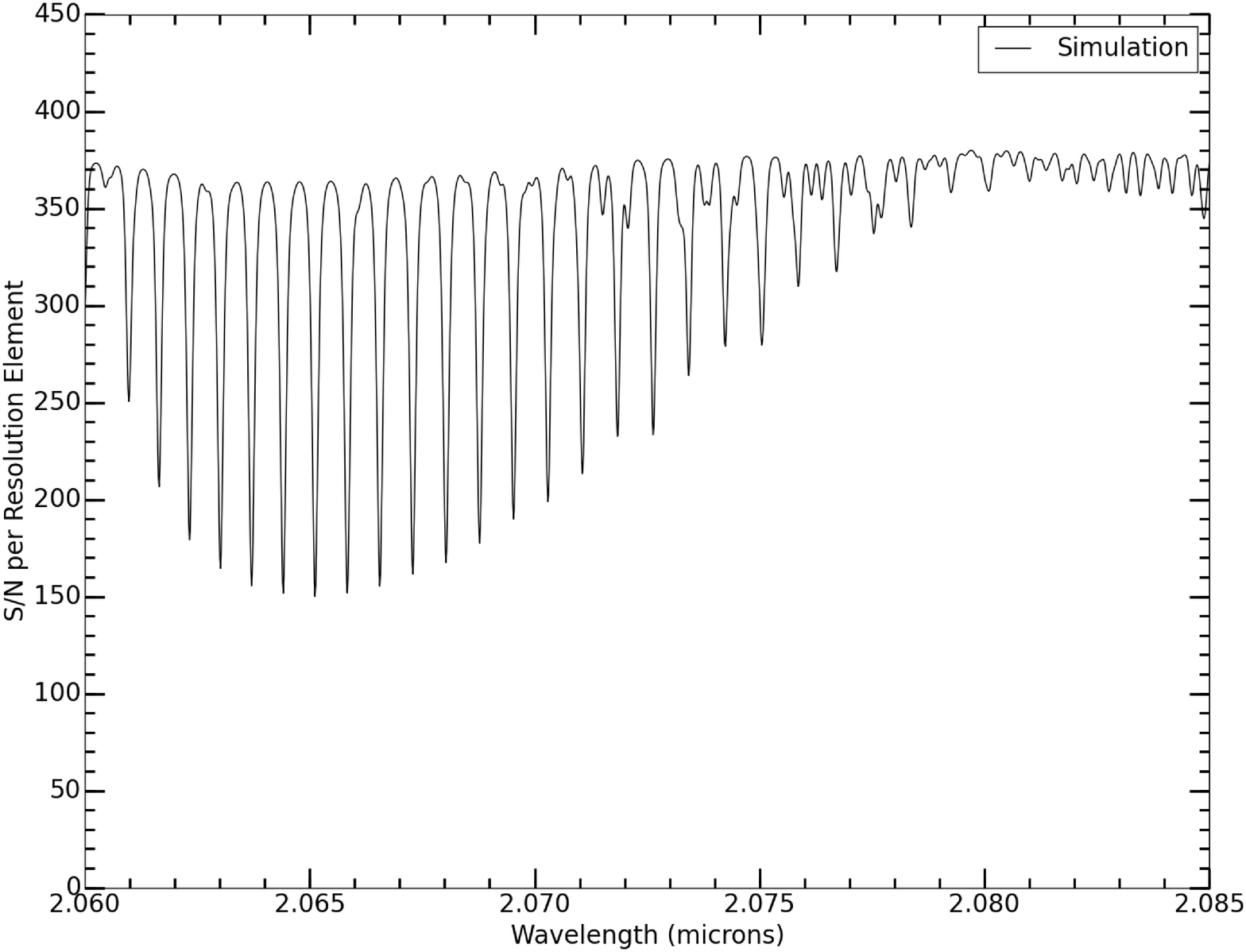}
\caption{Plots of S/N vs. wavelength of GSS~32 from the observational data (top plot) and simulation data from the ETC (bottom plot). The simulated data from the ETC assumes exposure times of t = 240 s, number of exposures, N = 4, the K-magnitude, m$_{K}$ = 7.324 mag, and the effective temperature of the source, T = 1360 K. The seeing is 0.8 arcsec. The PWV is assumed to be 2 mm. The Moon light is not included in the calculation.}
\label{snoise_gss32}
\end{figure*}

We compare the S/N of an early-type A0V star, HD~124683, and a late-type K5 star, GSS~32, to the S/N calculated using the ETC. We applied the same exposure-time, the number of exposures, K-magnitude, and the effective temperature of the sources in the ETC. We also applied the measured FWHM of the point spread function in the simulations. PWV is assumed to be 2 mm. Figure \ref{aov_sn} shows S/N of the real observational data and the simulated S/N of HD~124683. We estimate that the average S/N of the real observational data is 400, and the average value of the simulated S/N is 600. Then, the ratio of the simulated S/N and the real observational data S/N is 1.5, corresponding to 50$\%$. Figure \ref{snoise_gss32} shows the S/N values vs. wavelength of GSS~32. The average value of the simulated S/N and the real observational data S/N is 350 and 250, respectively. Then, the ratio is 1.4. The simulated S/N of GSS~32 is overestimated by 40$\%$.

The observed spectra of the stars are curved, while the simulated spectra are flat because we did not apply the blaze function to display the grating efficiency of echelle orders in the simulations. In the simulations, we focus on the calculations of the expected average S/N values, and simply compare that to the S/N values of the real observation data. The actual numbers would be different by a factor of 1.5 depending on the blaze angles.

\subsection{Signal-to-Noise for Emission Line}

\begin{figure*}[!h]
    \includegraphics[width=42ex,height=42ex,keepaspectratio]{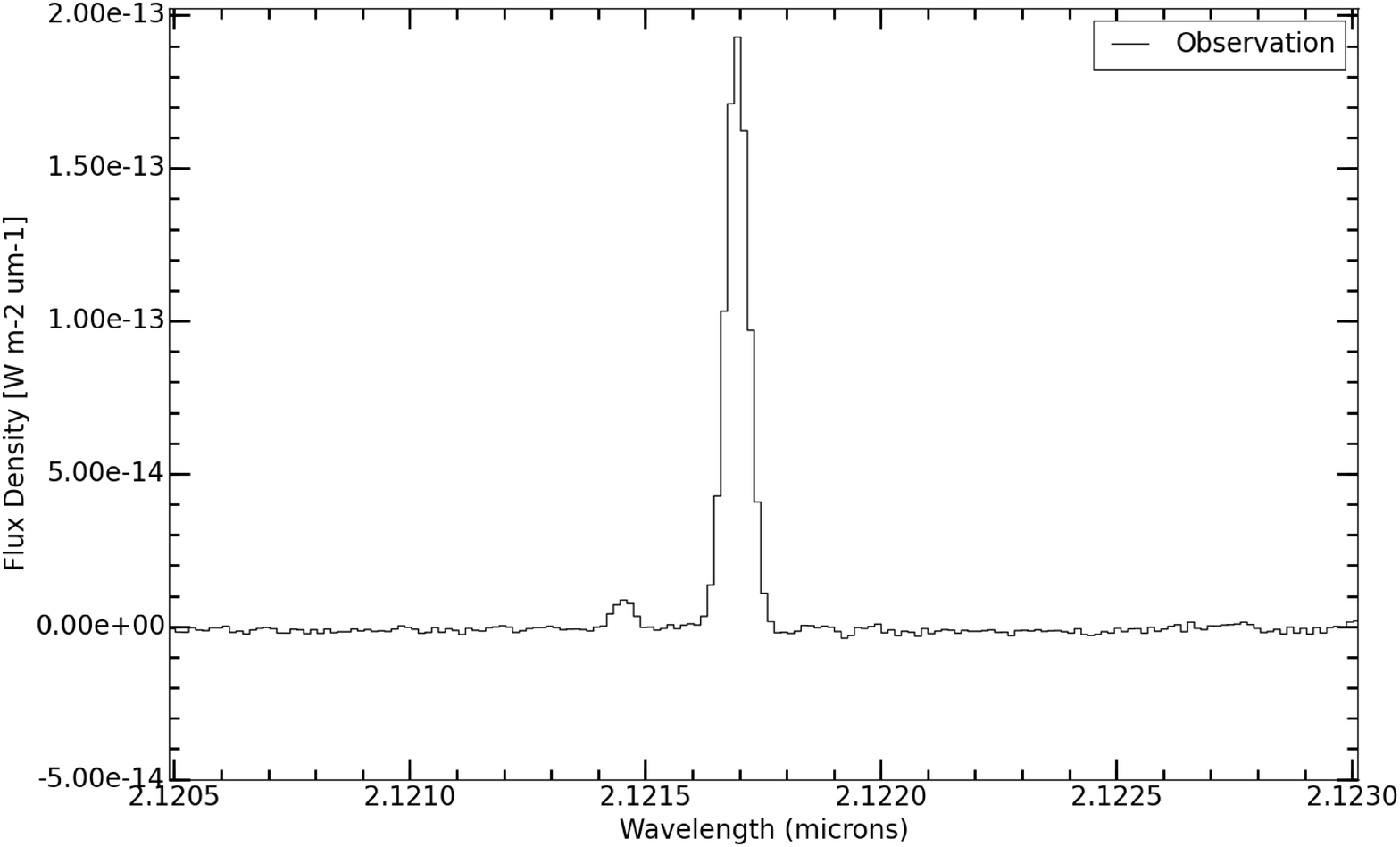}
    \includegraphics[width=42ex,height=42ex,keepaspectratio]{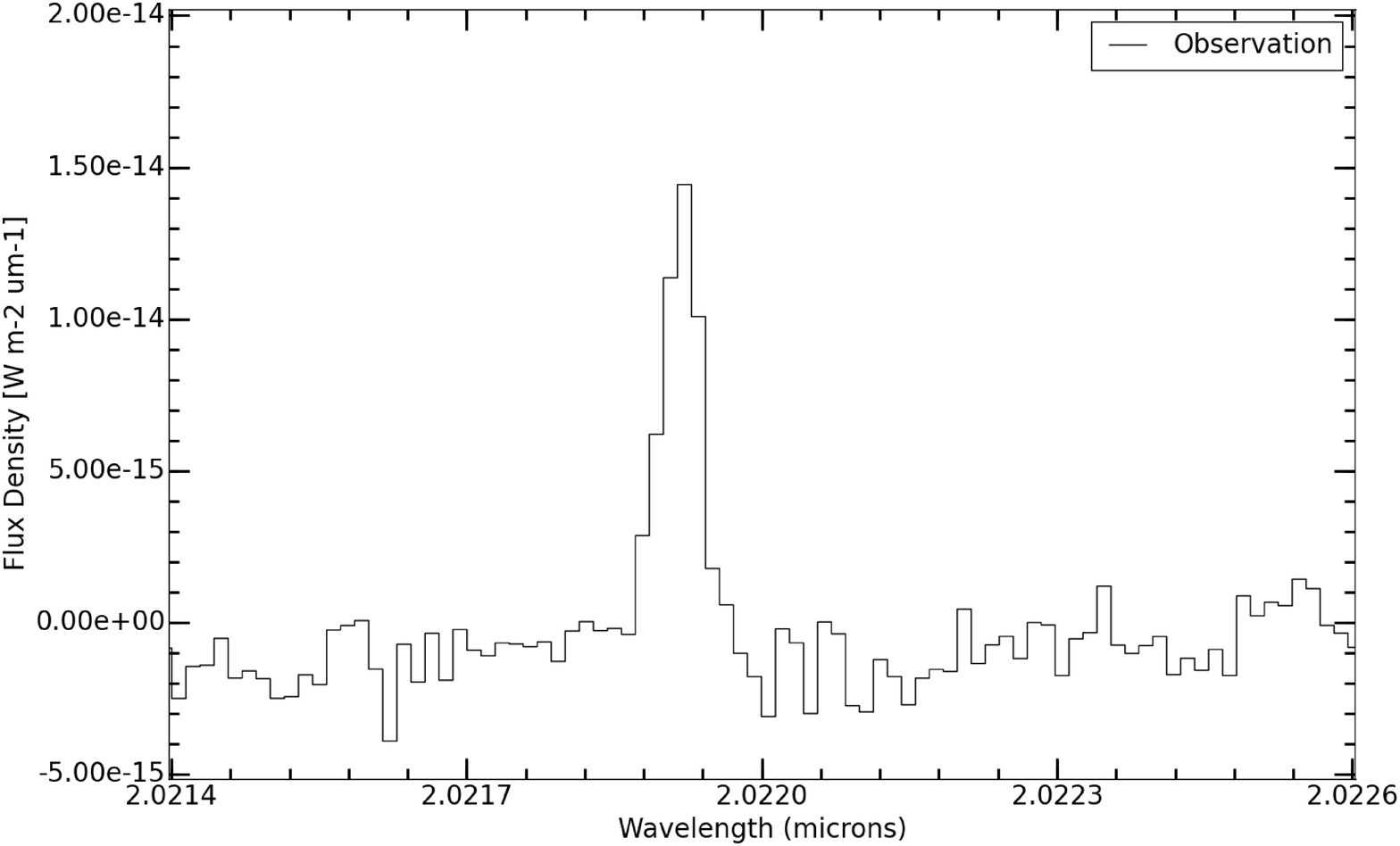}
    \hfill
    \includegraphics[width=42ex,height=42ex,keepaspectratio]{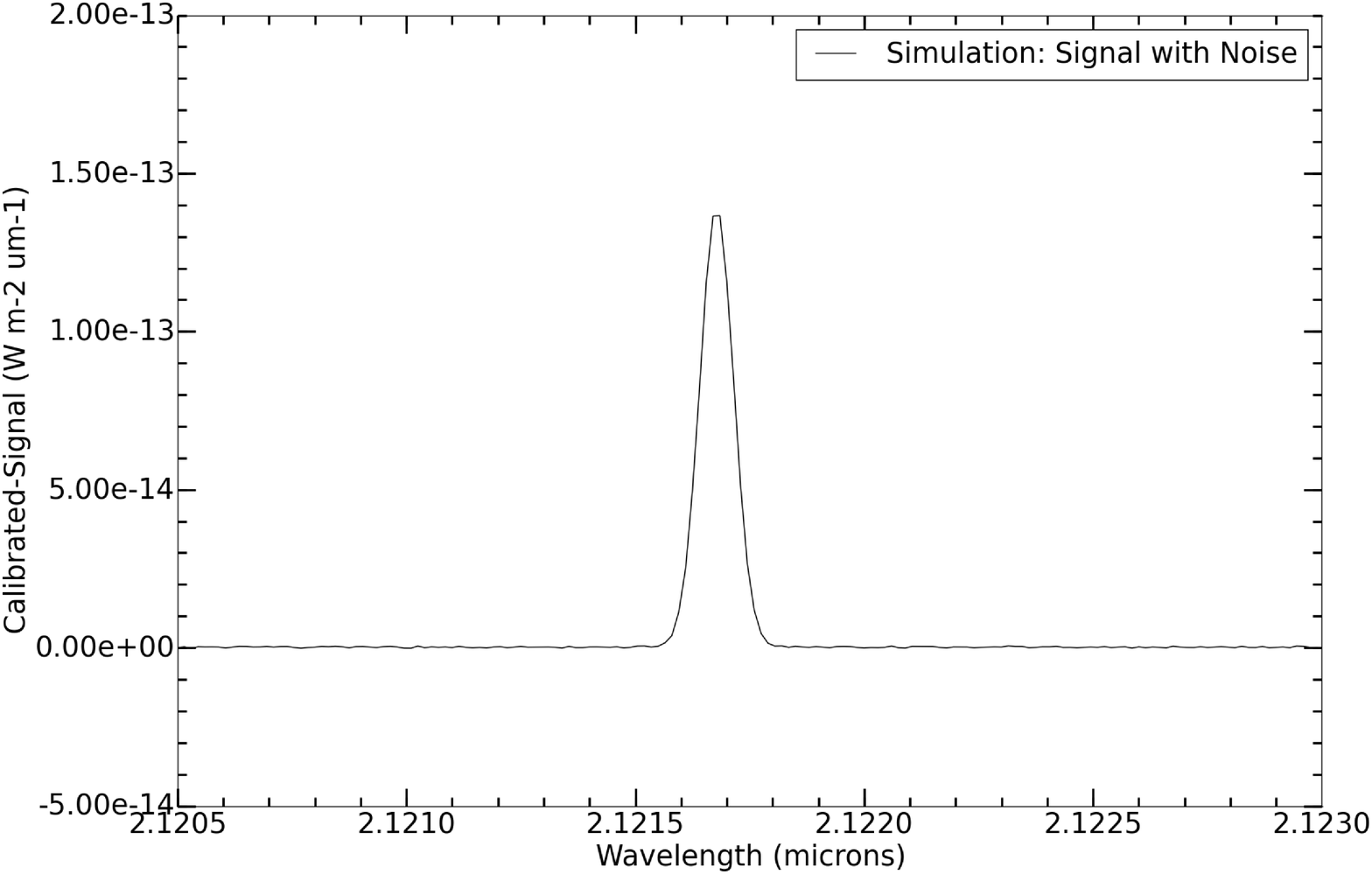}
    \includegraphics[width=42ex,height=42ex,keepaspectratio]{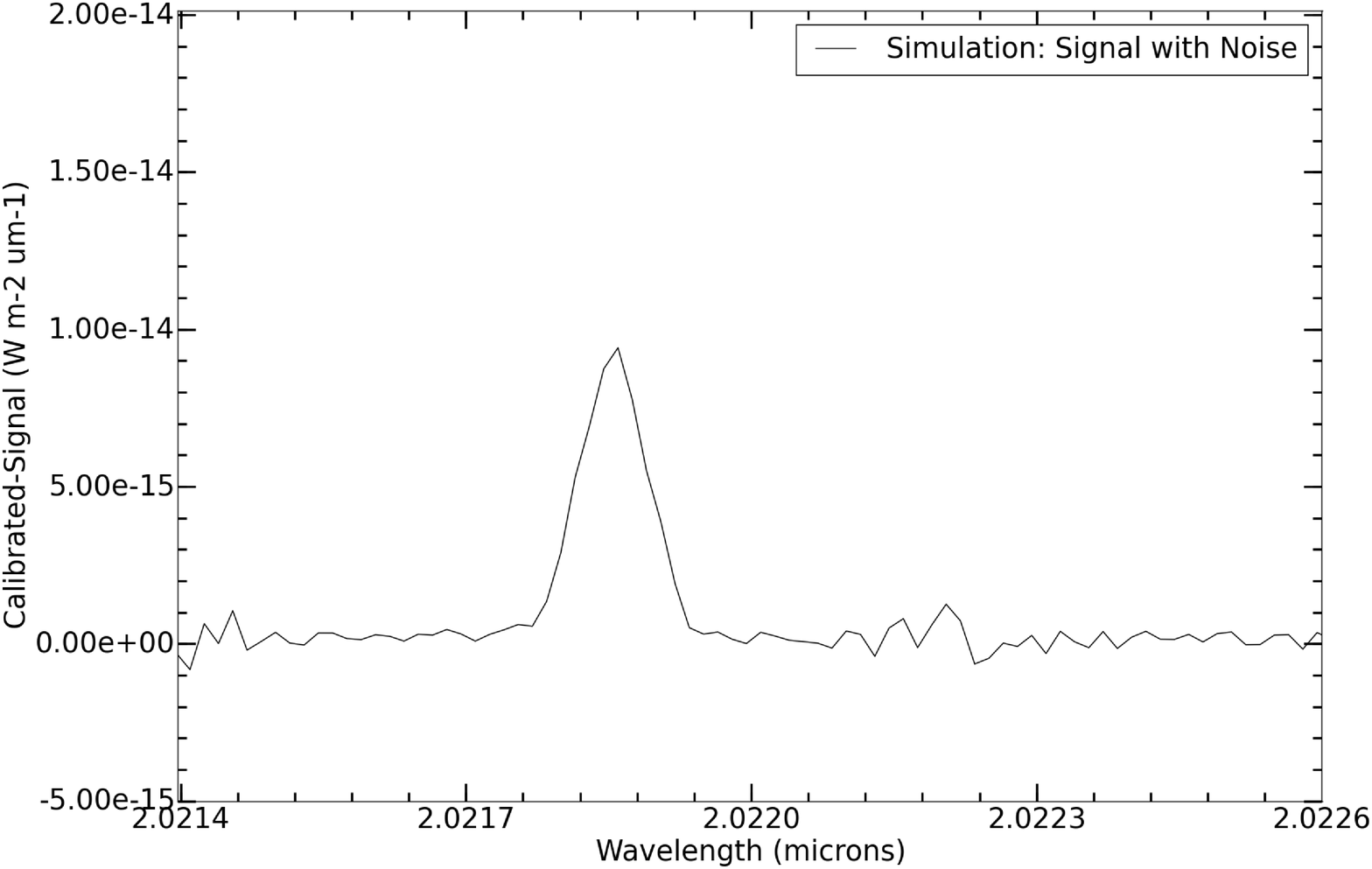}
\caption{Plots of observational data and the ETC simulation of NGC~7023. The left plots show the H$_{2}$~1-0~S(1) emission line, and the right plots show the H$_{2}$~7-0~O(5) emission line. We applied the Doppler shift corrections in the ETC, V$_{lsr}$ = -22.54 km s$^{-1}$, the local standard of rest velocity. The flux calibration is done by comparing to the signal from HD~155379. The Moon light is not included in the calculation.}
\label{h2}
\end{figure*}

We compared the simulated spectrum of NGC~7023 with those from the ETC. We applied the same exposure-time, the number of exposures, K-magnitude, and the effective temperature in the ETC. We chose a sample of the brightest emission line H$_{2}$~1-0~S(1) 2.12183 $\mu$m and the fainter emission line H$_{2}$~7-5~O(5) 2.02200 $\mu$m. In the ETC, we used the intensity values of the emission lines from \citet{Martini97}. The flux values which applied to the simulations of the IGRINS ETC are 12.8 $\times$ 10$^{-18}$ and 0.81 $\times$ 10$^{-18}$ W~m$^{-2}$ for H$_{2}$~1-0~S(1) and H$_{2}$~7-5~O(5), respectively. The line width value which inputted into the ETC is 9.4 km s$^{-1}$ based fitting the shape of the line profile using a Gaussian fitting method. We applied the Doppler shift correction in the ETC of V$_{lsr}$ = $-22.54$ km s$^{-1}$, this accounts for the local standard of rest radial velocity at the time of observation. Plots in Figure \ref{h2} show that the flux calibrations of the emission lines from the ETC and the observational data are approximate.

\section{SUMMARY}

We have developed the IGRINS ETC to generate realistic simulations of both continuum and emission line sources. The signal and noise values are estimated by taking into account telluric background emission and absorption, the emission and transmission of the telescope and instrument optics, and the dark current and read noise of the infrared detector arrays. We estimate the atmospheric transmission using models that account for the amount of precipitable water vapor (PWV) along the line of sight to the target. From the ETC, the observers can estimate the S/N of the spectrum for each spectral resolution element given the exposure-time and number of exposures.

The comparisons of the S/N vs. wavelength of HD~124683 and GSS~32 show that the simulated S/N values from the ETC are overestimated by 40$-$50$\%$. Note that the adopted throughput values in Table \ref{tau_emissivity} are mostly from rough estimates in ideal cases. In this paper, the comparisons of the simulated S/N values with that from the observed IGRINS data test the throughput dependent aspects of the ETC.

The source-code of the IGRINS ETC version 3.10 is available to be downloaded from the website http:$\slash$$\slash$irlab.khu.ac.kr$\slash$$\sim$igrins. \\

\section*{Acknowledgements}
This work was supported by the National Research Foundation of Korea (NRF) grant, No. 2008-0060544, funded by the Korean government (MSIP). This work used the Immersion Grating Infrared Spectrograph (IGRINS) that was developed under a collaboration between the University of Texas at Austin and the Korea Astronomy and Space Science Institute (KASI) with the financial support of the US National Science Foundation under grant AST$-$1229522, of the University of Texas at Austin, and of the Korean GMT Project of KASI. The IGRINS software Packages were developed based on contract between KASI and Kyung Hee University. We appreciate Dr. Wonseok Kang and Dr. Chae Kyung Sim for contributing the manual of the IGRINS ETC. We also thank Gordon Orris for proofreading this manuscript. This paper includes data taken at The McDonald Observatory of The University of Texas at Austin.

\begin{table*}[thp]
\caption{Parameters in throughput estimation}
\centering
\doublerulesep1.0pt
\renewcommand\arraystretch{1.5}
\begin{tabular}{ccc}
\hline
$\tau$              &  Parameter    & Value \\
\hline
$\tau_{\mathrm{atmo}}$              & Atmosphere$^{a}$ ($H$ band)    &  0.77   \\
$\tau_{\mathrm{atmo}}$              & Atmosphere$^{a}$ ($K$ band)    &  0.83   \\
$\tau_{\mathrm{mirror}}$            & Telescope Mirrors   &  0.55    \\
$\tau_{\mathrm{gr}}$  &  Immersion Grating    & 0.88           \\
$\tau_{\mathrm{vph}}$         & VPH Grating ($H$ band)  &  0.80   \\
$\tau_{\mathrm{vph}}$        & VPH Grating ($K$ band)  &  0.75   \\
$\tau_{\mathrm{optics}}$  & Other IGRINS Optics$^{b}$  &  0.42   \\
$\tau_{\mathrm{slit}}$        & Slit-Loss$^{a}$      &  0.64         \\
$\tau_{\mathrm{qe}}$  & Q.E. of Detector     & 0.80          \\[0.5ex]
\hline
\end{tabular}
\\
\scriptsize{$^{\rm a}$ See section 2.1.} \\
\scriptsize{$^{\rm b}$ Window, a pupil stop, collimators, and camera lenses.} \\
\label{tau_emissivity}
\end{table*}

\clearpage

\begin{table*}[thp]
\caption{Parameters in sensitivity simulations}
\centering
\doublerulesep1.0pt
\renewcommand\arraystretch{1.5}
\resizebox{15cm}{!}{
    \begin{tabular}{cccc}
    \hline
    Parameter                               & $H$ band            & $K$ band                  & Unit      \\
    \hline
    Center Wavelength                                      &  1.63       &  2.22                       &   $\mu$m                      \\
    Zodiacal Light$^{\rm a}$ at latitude of 0$^{\circ}$             &  99   &  73                             & photon s$^{-1}$ m$^{-2}$ arcsec$^{-2}$                   \\
    Zodiacal Light$^{\rm a}$ at latitude of 30$^{\circ}$              &  50  & 37                                &  photon s$^{-1}$ m$^{-2}$ arcsec$^{-2}$                \\
    Zodiacal Light$^{\rm a}$ at latitude of 60$^{\circ}$              &  42  & 31                                &  photon s$^{-1}$ m$^{-2}$ arcsec$^{-2}$                \\
    Zodiacal Light$^{\rm a}$ at latitude of 90$^{\circ}$              &  30  & 22                                &   photon s$^{-1}$ m$^{-2}$ arcsec$^{-2}$               \\
    Half Moon light$^{\rm b}$                               &  54 & 30                    &   photon s$^{-1}$ m$^{-2}$ arcsec$^{-2}$              \\
    Full Moon light$^{\rm b}$                               &  595  & 277                       &  photon s$^{-1}$ m$^{-2}$ arcsec$^{-2}$              \\
    OH light$^{\rm c}$                                 &   3390   & 1096                    &  photon s$^{-1}$ m$^{-2}$ arcsec$^{-2}$      \\
    Ambient Temperature                     &   275       &  275                                &    K  \\
    Detector Read Noise                     &    5   &    5                        &    electron     \\
    Detector Dark Current                   &   0.02  &   0.02    &   electron~s$^{-1}$     \\ [0.5ex]
    \hline
    \end{tabular}
    }
    \\
    \scriptsize{$^{\rm a}$ See section 2.2. We assume the heliocentric ecliptic longitude is $\pm$90$^{\circ}$.} \\
    \scriptsize{$^{\rm b}$ See section 2.2. We assume the separation between target and Moon is 60$^{\circ}$, and the altitude of the Moon is 45$^{\circ}$.} \\
    \scriptsize{$^{\rm c}$ See section 2.2. The OH value use in equation (\ref{eq:zod}) is total OH line intensity in $H$ and $K$ bands.} \\
\label{background}
\end{table*}

\clearpage

\begin{table*}[thp]
\caption{Average intensity values in Flat image}
\centering
\doublerulesep1.0pt
\renewcommand\arraystretch{1.5}
\begin{tabular}{cccc}
\hline
Detector            & Order regions          & Inter-order regions    &   Ratio \\
\hline
$H$ band               &  4208               &  698             &   0.20$^{\rm a}$ \\
$K$ band               &  2345               &  650             &   0.38$^{\rm a}$  \\ [0.5ex]
\hline
\end{tabular}\\
\scriptsize{$^{\rm a}$ We calculate the ratio for both $H$ and $K$ bands by [(698+650)/2]/[(4208+2345)-(698+650)/2] = 0.1.} \\
\label{order}
\end{table*}

\clearpage

%


\begin{table*}[thp]
\caption{Standard Stars}
\centering
\doublerulesep1.0pt
\renewcommand\arraystretch{1.5}
\begin{tabular}{cccccc}
\hline
Standard star & Observing date (UT) & $t_{exp}$  & $m_{K}$$^{a}$  &  Type$^{a}$  & $t_{eff}$$^{a}$  \\
                &                   &   (s)  &  (mag)     &         &    (K)           \\
\hline
HD~124683     &    2014 May 26  & 120  &  5.552  &  A0V  &  9500 K     \\
HD~155379    &    2014 Jul 12  & 120  & 6.520  &  A0V  &  9500 K    \\ [0.5ex]
\hline
\end{tabular}
\\
\scriptsize{$^{\rm a}$ http:$\slash$$\slash$simbad.u-strasbg.fr$\slash$simbad$\slash$sim-fbasic.}
\label{standard}
\end{table*}

\clearpage


\clearpage


\clearpage

\clearpage

\clearpage

\clearpage


%

\clearpage

\appendix

\section{MODEL ATMOSPHERE}
\begin{sidewaysfigure}[!thp]
\centering
    \includegraphics[width=120ex]{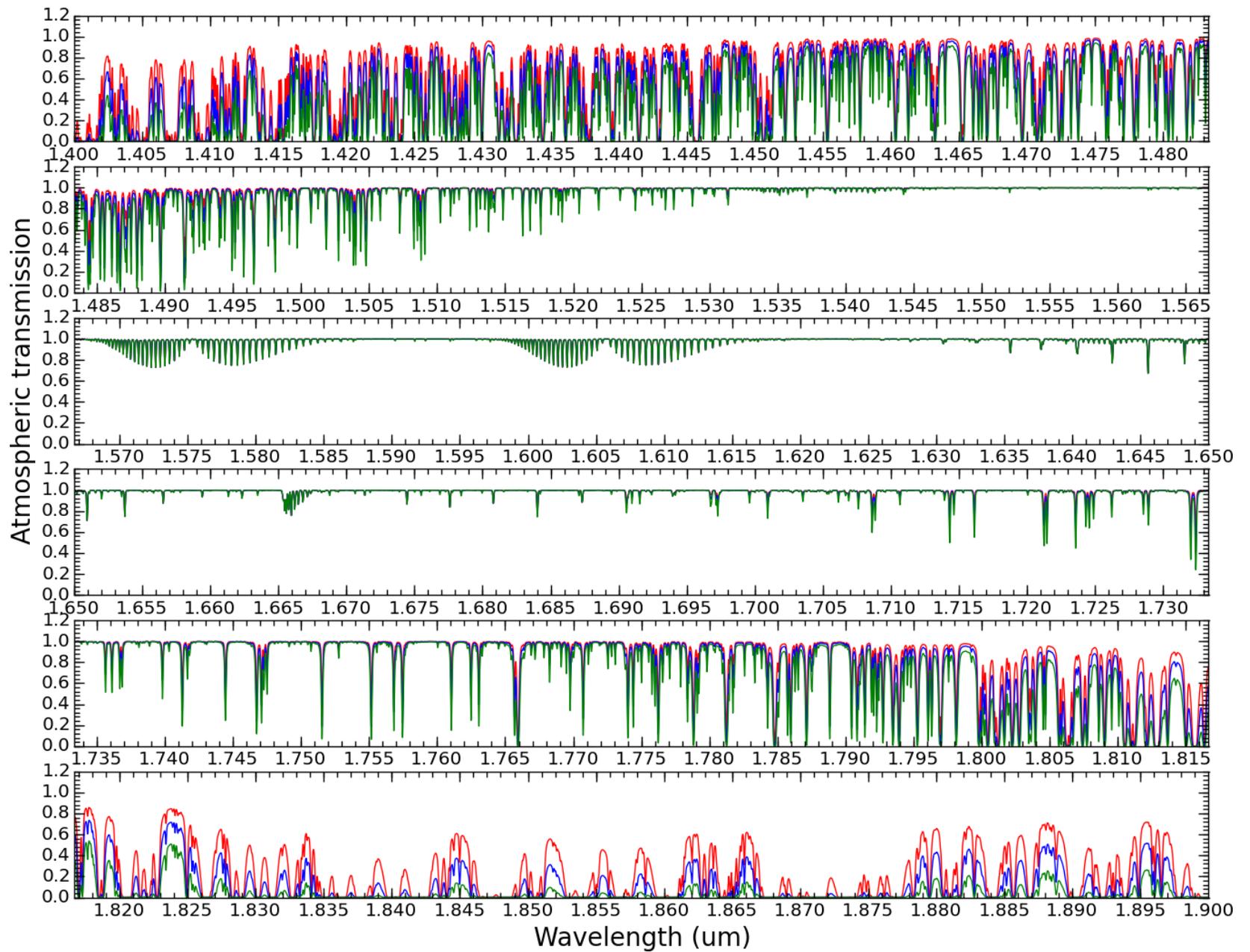}
\caption{Plots of atmospheric transmission in the $H$ band. Precipitable water vapor (PWV) of 2 mm (red lines), 4 mm (blue lines), and 8 mm (green lines) for the weather conditions in the McDonald Observatory.}
\label{atmospheric_hband}
\end{sidewaysfigure}

\begin{sidewaysfigure}[!thp]
\centering
    \includegraphics[width=120ex]{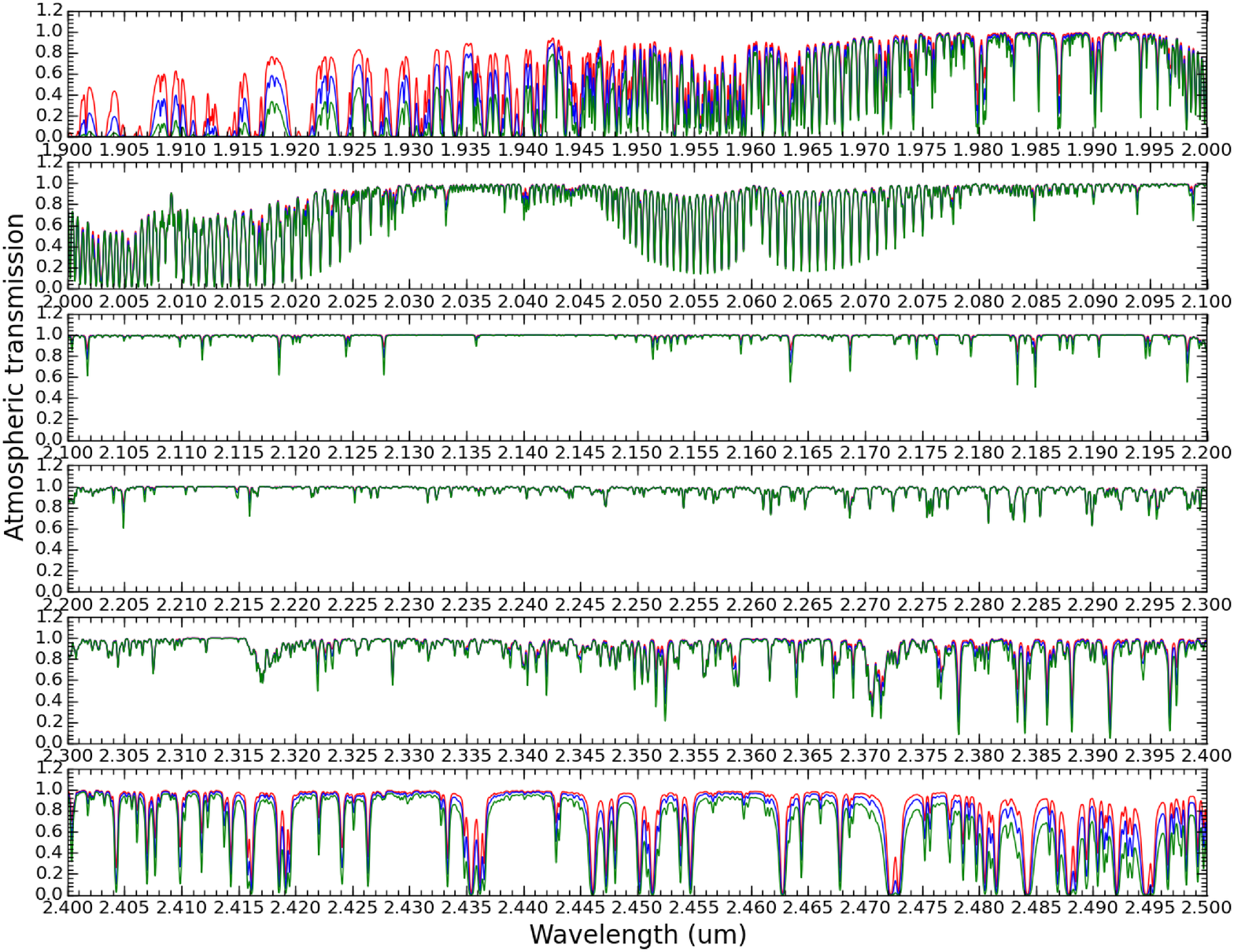}
\caption{Atmospheric transmission in the $K$ band. Precipitable water vapor (PWV) of 2 mm (red lines), 4 mm (blue lines), and 8 mm (green lines) for the weather conditions in the McDonald Observatory.}
\label{atmospheric_kband}
\end{sidewaysfigure}

We consider telluric absorption and emission lines in the IGRINS ETC using a model to calculate synthetic telluric spectra \citep{Seifahrt10}. The model spectra cover from 1.4$-$2.5 $\mu$m with precipitable water vapor (PWV) of 2, 4, and 8 mm for the weather conditions at McDonald Observatory. The original transmission atmosphere spectra are convolved to be the same spectral resolution as IGRINS, $R$ $=$ 40,000 in the $H$ and $K$ bands. Figure \ref{atmospheric_hband} and \ref{atmospheric_kband} show the atmospheric transmission in the $H$ and $K$ bands.

\begin{figure*}[p]
\centering
    \includegraphics[width=60ex,height=60ex,keepaspectratio]{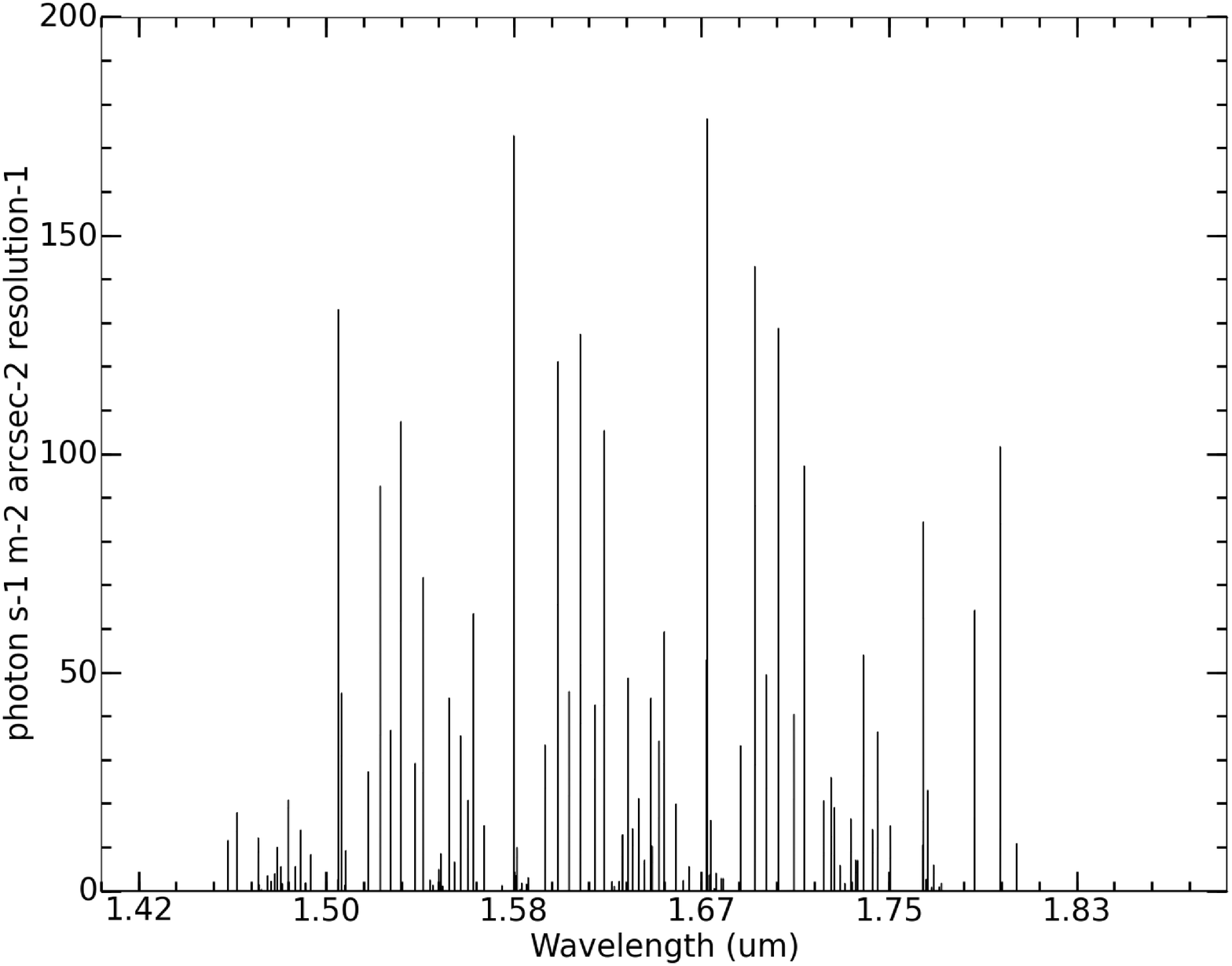}
    \includegraphics[width=62ex,height=62ex,keepaspectratio]{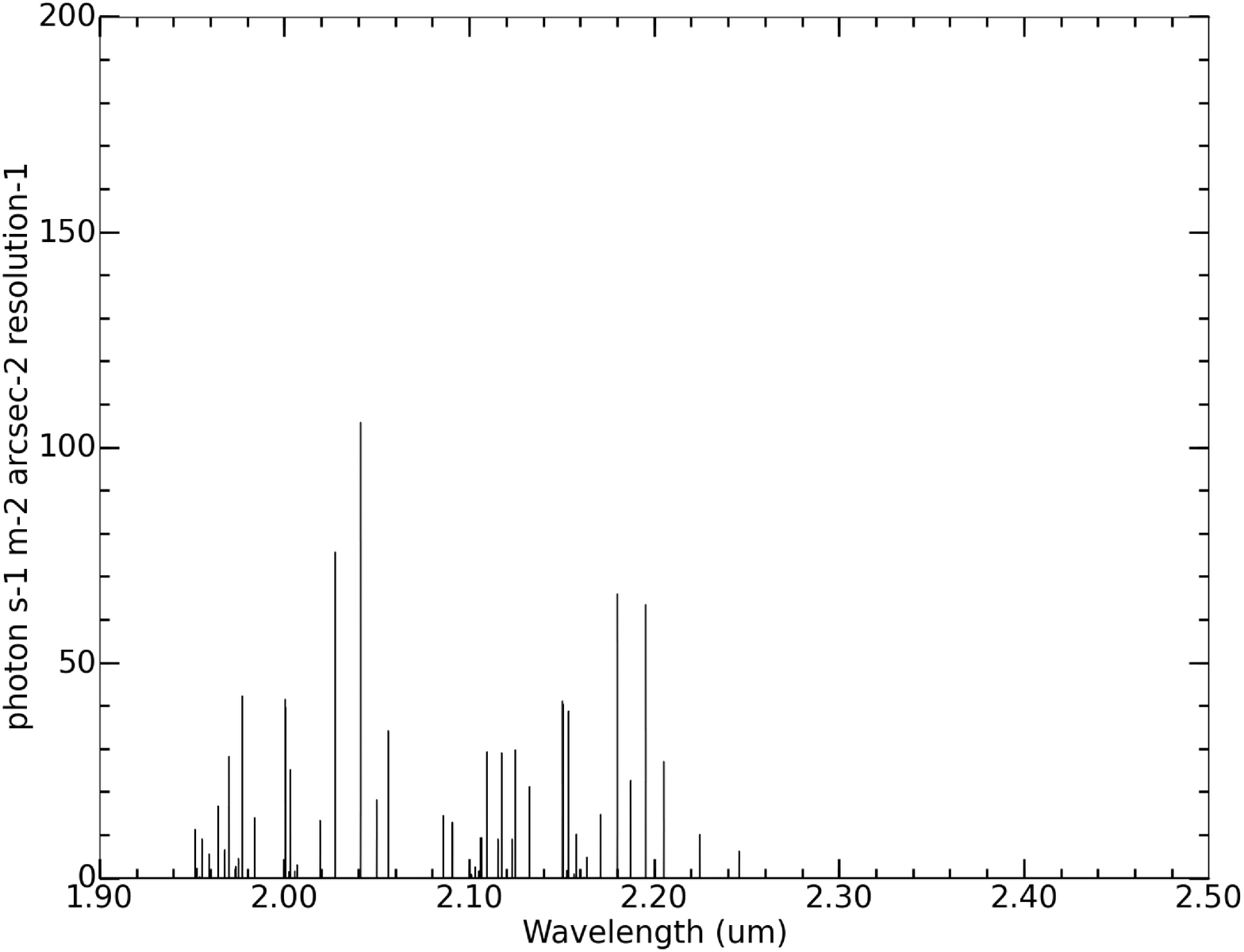}
\caption{Plots of OH emission lines in the $H$ band (top plot) and $K$ band (bottom plot).}
\label{ohlines}
\end{figure*}

We compiled lists of 249 bright OH emission lines from the literatures (\citealp{Rousselot00}; \citealp{Oliva92}). The original OH emission line data are convolved to be the resolution of $R$ $=$ 40,000. Figure \ref{ohlines} shows the spectra of OH emission lines in the $H$ band and $K$ band.

\clearpage

\clearpage


\clearpage

\end{document}